\documentclass[12pt,a4paper]{article}
\usepackage{graphicx}
\usepackage{amsfonts}
\begin{document}
\title{Introduction of coherence in astrophysical spectroscopy}
\author{Jacques Moret-Bailly.\\
 \it Mail: jmo@laposte.net}

\maketitle

\begin{abstract}
By confusing the radiance of a single mode light beam, constant in a transparent medium, with the irradiance which decreases away from the source, Menzel purports to show that coherent interactions of light with the diluted media of astrophysics, are negligible. Therefore, to study the interaction of light with gases, astrophysicists use Monte Carlo computations which work to study nuclear systems, but not optics: optical modes which may be defined in inhomogeneous media or for the emissions of single atoms interact coherently with these systems: a unique formula represents, according to the sign of a parameter, absorption and coherent emission. The optical and spectroscopic properties of a very simple model, an extremely hot source in an isotropic cloud of pure, low pressure, initially cold, huge hydrogen cloud are studied using Planck's and Einstein's theories. The similarities of the images and the spectra of this simple model with astronomical observations, for instance of SNR1987A, Einstein cross, lyman break galaxies, quasars,... is so large that this model may be an elementary first step in the study of many astrophysical objects. Adaptations of the model to complex astrophysical systems could represent them using only the old, standard theories of physics commonly used in laser spectroscopy.

\end{abstract}
Keywords: {Radiative transfer; ISM:H$_{II}$ regions.}
{010.5620;030.1640;190.0190}

\section{Introduction}
Referring to Menzel \cite{Menzel} who confused radiance and irradiance, many astrophysicists believe that coherent interactions are negligible in astrophysical systems. The best physicists have long had similar ruling, stating for example: \textquotedblleft Townes' maser will not work\textquotedblright. While the development of experiments using lasers has led spectroscopists to study the coherent interactions between light and matter, astrophysicists have kept their wrong methods, especially using Monte-Carlo calculations apparently inspired by nuclear physics. 
Deducting mainly results from thermodynamics, Planck and Einstein have obtained very general laws, independent for instance of the use of a classical or quantum theories of light. A unique formula represents, according to the sign of the amplification coefficient of the radiance of a light beam, absorption and coherent amplification.

Section \ref{optics} recalls general concepts of coherent optics. This is not useless, recalling, for instance how Planck's and Einstein's theories must replace the Monte-Carlo method to compute the propagation of light in a resonant medium, and that some interpretations of photon-counting signals are wrong.

Section \ref{utile} describes optical effects well known in laser spectroscopy, which seem apply easily in astrophysics. It shows that usual Monte-Carlo computations are wrong in the low density media of astrophysics.

Section \ref{possible} Proposes to apply these optical effects to find simple explanations of many observations.

\section{Reminding general concepts of optics.}\label{optics}
\subsection{The optical modes.}

Optical modes are needed to study classical and quantum theories of light.

Maxwell's equations are linear in vacuum, so any linear combination of solutions is a solution. Thus, solutions are represented by points in a vector space $\mathfrak{S}$. We call \textquotedblleft mode \textquotedblright{} a set of solutions that differ only by a real multiplicative factor, that is a radius of $\mathfrak{S}$. The norm of an electromagnetic field in a space {\it empty of any other field} is the energy of this field. Two fields (consequently two modes) are orthogonal if the norm of their sum is the sum of the norms of each field. For theory, matter is not introduced by permeability and permittivity, but by the elementary sources (electrons, nucleus) which, following Schwarzschild and Fokker are replaced by their advanced fields, so that the field equations remain linear and sets of modes may be defined.

 An usual type of mode is defined from a nearly monochromatic, polarized electromagnetic beam limited by two circular holes and diffraction. Its flux of energy by unit of frequency is its spectral radiance\footnote{Using only spectral radiances in this paper, \textquotedblleft radiance\textquotedblright{}  will mean \textquotedblleft spectral radiance\textquotedblright.}. The modes deduced from the general definition may be extremely complex: their frequency band is not necessarily sharp, they may be polychromatic; their geometry may be complex, for instance tortuous in opalescent media. Thus, some useful general concepts, for instance radiance, may require new definitions. We will remind the general concepts using usual modes, else results remain valid while it may be difficult to set their equations.

\subsection{Light-matter interactions; Einstein B coefficient.}
Light may diverge far from a source into a very large region. Thus coherent polarization by a single-mode light beam may concern all atoms in a large volume. These atoms make a single polarized quantum system which may interact with a single photon. In this large quantum system, in a first approximation close to the classical interpretation, any atom is considered as \textquotedblleft dressed\textquotedblright{} by the field, its stationary state being slightly mixed with its other states. Polarization can absorb a significant energy to be returned to one or several exciting wave(s) when light disappears because the atoms must return to their original state.

Einstein introduced by thermodynamics the theory of propagation of light in a resonant medium, theory which uses A and B coefficients \cite{Einstein1917}. In a large medium, homogeneous enough to use standard optics, the coherent amplification corresponding to B may be easily connected to Huygens' theory of the propagation of monochromatic waves: Huygens construction uses wavelets emitted by all points of a wave surface $W$ to generate an other wave surface $W_1$, envelope of the wavelets. The retrograde wave is eliminated by the radiation of a wave surface whose distance from $W$ is a fourth of wave. Suppose that light propagates in a gas such that the number of molecules interacting in an homogeneous region is large. Wavelets generated by molecules lying on $W$, coherently (that is with an identical phase), at the same frequency (Rayleigh scattering) generate also $W_1$. Thus, the amplitudes of the incident and scattered waves may be added, the result being a single wave, with the same frequency. For simpler computations, the scattered field is split into two fields: - a field delayed by $\pi/2$ which produces the refraction; - a field having the same or an opposite phase which produces a positive or negative amplification of the incident wave. The scattered amplitude being supposed proportional to the incident amplitude, the amplification of the amplitude (or of the radiance) is proportional to the incident amplitude (or radiance). Refraction changes the geometry of the modes while amplification does not. 

Einstein's theory, restricted to coherent amplification of a light beam propagating in an homogeneous medium, and Huygens' theory have the same limits of validity based on statistics: Between two wave surfaces distant of a wavelength, matter must be homogeneous and the number of molecules having the same properties must be large enough to apply the law of large numbers.

Einstein theory applies to all modes. Choosing a particular electric charge, it applies to the advanced and retarded modes of absorption and emission of this charge. The theory of amplification may become: the ratio of the amplitudes of the advanced and retarded fields is a real number. To introduce also refraction of quasi-monochromatic light, this number is set complex.

\medskip
The energy exchanged with matter is quantized. In classical and semi-classical representations, the energy in a mode may vary continuously. In quantum representation of light, the modes have the energies of a quantum oscillator, so that the energy of a mode jumps between levels having exactly energies $(n+1/2)h\nu$, with n integer.

\subsection{Relative and absolute radiances; Einstein A coefficient.}
Atomic theory shows that the distances between elementary charges (electrons, nuclei) are much larger than the size of these charges, so the field emitted by a charge $q$ in its specific mode is much larger in its neighborhood than the fields created by other charges. An opposite field is needed to absorb the field\footnote{This trivial and rigorous conception of absorption of a field explains, for example, the dynamics of starting up of lasers, avoiding the introduction of ad hoc concepts as radiation reaction.} emitted by $q$. The generation of this opposite field, in particular in the vicinity of $q$, requires fields created by numerous other charges: The absorption of all fields is impossible, beyond an absorber it remains a {\it residual spectral radiance}\footnote{The residual field or radiance have many other qualifiers: {\it zero point; minimal}; {\it stochastic}, while it is only stochastic far from the sources; {\it quantum} while its existence was found before quantum mechanics,...} whose mean value $h\nu^3/c^2$ in a monomode beam results from the computation of absolute energies needed, at least theoretically, to apply formula $E=mc^2$ \cite{Planck1911,Einstein1913}. Consequently the absolute spectral radiance of a monomode beam in a blackbody is $I_\nu=(h\nu^3/c^2)\{1+2/[exp(h\nu/kT_{P\nu})-1]\}$. Subtracting  $h\nu^3/c^2$ from an absolute field gives the usual, particular relative field set equal to 0 in the dark. Planck's formula allows to associate, outside of a blackbody, a radiant Planck's temperature $T_{P\nu}$ to a region of radiance $I_\nu$ of a monomode beam.

{\it Radiance is the fundamental parameter characterizing the flux of energy in a progressive light mode. Spectral radiance allows to associate a temperature to this flux, then to compute entropy.}

Using the absolute field, Einstein's A coefficient is null, so that the theory is simple, therefore reliable: A monomode beam coming from an unknown dark source, has an unknown, therefore stochastic phase, and its spectral radiance is, in the average, $h\nu^3/c^2$. Clausius showed that, in a transparent medium, the radiance is constant along a beam; in matter, the amplification of radiance depends on $B$ coefficient. Let $F=(1+\alpha)Z$ the absolute field resulting from amplification of a residual field $Z$, of {\it average} radiance $h\nu^3/c^2$ and $f=F-Z=\alpha Z$, the usual relative field. An opaque photocell absorbs a quantified flux of energy whose mean value $E$ is proportional to $F^2-Z^2=2FZ+f^2$ because it remains the field $Z$ in the mode. Using a relative radiance, it is often written that $E$ is proportional to $f^2$, even in photon counting while $2FZ$ is predominant\footnote{This error allows to demonstrate the superiority of quantum electrodynamics on classic electrodynamics in the experiments using photon countings. Bell inequalities verify the incompatibility of Copenhagen postulates with classic theories.}. If two monomode beams are non orthogonal to an initial monomode beam, fluctuations of $Z$ in the initial beam are involved in the observed beams where they are correlated (photon bunching). Observations of these correlated fluctuations may come from a space cutting of light of a star by a telescope or from a time cutting.

\subsection{Monte-Carlo computations.}
The name \textquotedblleft Monte-Carlo\textquotedblright{} means that a random variable is introduced in a calculation, then that the results are averaged. These computations can be useful when light propagates in a medium too inhomogeneous (eg opalescent near the critical point of a gas) for surface wave of radiation can be defined. But the need of this introduction is necessary only because a too complex system cannot be fully known: Einstein theory rules fundamentally the light-matter interactions.
At medium pressure, density fluctuations are responsible for scatterings that are incoherent, for example, the blue sky. These scatterings are induced by the residual fields of the emission modes of the fluctuations and take their phases. These phases are usually unknown, but photon bunchings are possible. At low pressure, the fluctuations are produced mainly by binary collisions whose density is proportional to the square of the pressure.

While thermodynamics provides only average values, its use is essential to the interpretation of macroscopic observations. This involves using radiance of light to get its temperature. In the interstellar medium, the gas density is too low for collisions produce a significant incoherent scattering. Thus, using a Monte Carlo computation is not justified.

\section{Coherent light-matter interactions useful in astrophysics.}\label{utile}
\subsection{Superradiance and competition of modes.}

Boltzman temperature $T_{B\nu}$ of a set of atoms for a transition of frequency $\nu$ corresponds to an equality between $exp(h\nu/kT_{B\nu})$ and the ratio of populations of atoms in the upper and lower levels of the transition (with this definition, $T_{B\nu}$ may be negative).
A positive amplification of a single mode beam can produce a large radiance, if $T_{B\nu}>0$ up to a limit corresponding to $T_{B\nu}=T_{P\nu}$, without limit if $T_{B\nu}<0$. If the radiance becomes larger than an arbitrary limit, for instance $2h\nu^3/c^2$, the ray is called \textquotedblleft superradiant\textquotedblright.

The gas being isotropic, if the temperatures $T_P$ of two beams at their intersection are much lower than $T_B$, or if $T_B<0$, their amplification coefficients have nearly the same value. Since the exchanged energy is proportional to the incident energy, the more powerful beam exchanges more energy. For a positive amplification, it can absorb the bulk of available energy: This \textquotedblleft competition of modes\textquotedblright{} leaves only a few modes get a high radiance.

\subsection{Multiphotonic and parametric interactions.}

Dressed by several rays of light, atoms can undergo a series of transitions. If the radiances of the rays are large, the atoms pass directly from their initial state to their final state: the interaction is called \textquotedblleft multiphotonic\textquotedblright. Few intermediate states may be non-stationary, virtual.

A set of multiphotonic interactions that returns the atoms to their original states, is a \textquotedblleft  parametric interaction\textquotedblright, where the atoms act as a catalyst. It is obviously an increase in the entropy of light rays, computed from Planck's law. Several types of atoms can participate in a parametric effect which combines the frequencies, but each atom must complete a full cycle of transitions. The parametric interactions are especially intense because atoms do not need to acquire an energy that would enable them to reach another stationary state: their initial stationary state is only slightly mixed with other states during polarization.

\subsection{Impulsive Raman scattering.}

The coherent Rayleigh scattering that produces the refraction can be replaced by a Raman scattering if the interaction remains parametric and if the conditions of coherence written by G. L. Lamb \cite{Lamb} are satisfied: the time-coherence (the length of the pulses which model the time-incoherent light) must be \textquotedblleft shorter than all implied time constants\textquotedblright{}. These constants are the period of a Raman resonance and the collisional time.

With Lamb's conditions, the spectral lines of exciting and scattered beams overlap. An elementary calculation shows that, for an interaction in an infinitesimal layer of gas, these beams interfere into a single monochromatic beam whose frequency is intermediate, in proportion to the amplitudes. The infinitesimal shifts of frequency add along the path of the ray. To obtain a parametric interaction, the energy transferred to matter by the Raman scattering must be compensated by at least an other simultaneous, coherent Raman scattering.

The \textquotedblleft Impulsive Stimulated Raman Scattering\textquotedblright{} uses two beams of femtosecond, powerful laser pulses. It is a nonlinear interaction easily observed in the laboratory  \cite{Weiner}. By using a single lower power, pulsed laser, the experiment requires a much larger length of interaction. The effect appears in long optical fibers used in telecommunications: the pulse frequency is reduced proportionally to the length of the fiber, energy is transferred to thermal radiation. This effect complicates the multiplexing of information transmitted at several colors by long fibers. The frequency shift is, in a first approximation, inversely proportional to the cube of the length of pulses \cite{MB}. Natural light is made of pulses of about 1 nanosecond, $10^5$ times longer than the femtosecond laser pulses. Thus, the observation of a frequency shift by a Coherent Raman Effect on Incoherent Light (CREIL) requires an interaction length $10^{15}$ times longer than the length of used optical fibers, length not available in a laboratory.

With natural light, Lamb's conditions impose a  collisional time larger than 1 nanosecond, so the use of a low pressure gas. In this gas, there must be a quadrupolar resonance whose period is larger than 1 ns: The resonance at 1420 MHz of the hydrogen atom is not appropriate, but the resonance frequencies 178 MHz in the 2S$_{1/2}$ state, 59 MHz in 2P$_{1/2}$ state, and 24 MHz in 2P$_{3/2}$ are suitable.

\section{Possible applications in astrophysics.}\label{possible}
\subsection{Superradiance and competition of modes.}

Str\"omgren \cite{Stromgren} studied a system composed of an extremely hot source (temperature larger than $3.10^5 K$) immersed in a cloud of hydrogen at low pressure, initially cold. He showed that where the temperature becomes low enough that neutral hydrogen atoms are formed, their intense radiation cools the gas so that it forms a relatively thin shell (Str\"omgren shell) of neutral excited atomic hydrogen surrounding an almost completely ionized sphere (Str\"omgren sphere).
The intense superradiance was little known in 1939, so Str\"omgren did not introduce a competition of modes. Let us study the amplification of rays that pass through the system at a distance $r$ from the source. Two rays L$_1$ and L$_2$ passing at weak distances $r_1$ and $r_2$ from the source cross twice the shell. Str\"omgren's shell can be divided into infinitesimal concentric shells. If $r_1<r_2$, the path of radius L$_1$ in each infinitesimal shell is shorter than the path of L$_2$, thus L$_1$ is less amplified than L$_2$, amplification is thus an increasing function of $r$. As the amplification is zero if the ray does not penetrate the shell, the amplification function has at least a maximum value for a value $R$ of r (The minimal R is chosen if there are several maxima).

The brightest rays are tangent to the sphere of radius $R$. Into a given direction, these rays make a cylinder and the system is seen as a ring. If the superradiance is very intense, there is a competition of the modes on the cylinder, if a spectral line is selected the ring is dotted. It has the appearance of TEM$_{p,l}$ modes of a laser selected for a fixed radial order $p$ and the $2(p+1)$ values of angular index $l$. 
The rings may appear for several transitions, but these rings and dots are not independent: if a superradiance is very intense, for atoms located on the path of the beam, the upper state of the transition is depleted while the lower is more populated. This increases the intensity of an other transition which arrives at the upper state, or starts at the lower. The dots at long wavelengths are large, the dots at shorter wavelengths appear at their inner rim.

\medskip
This explanation is simpler than the usual explanations of the \textquotedblleft pearl necklaces\textquotedblright{} of SNR1987A, of Einstein cross or arcs.

In the standard explanation of pearl necklace SNR1987A, a shock wave created in H$_{II}$ by the explosion of the star hits a ring of relatively dense H$_{I}$ that is excited by the collision \cite{1987A,Chevalier}. The propagation time of the shock wave explains that the ring does not appear immediately after the explosion, but the persistence of its brightness is amazing. For much of the energy of the shock wave is not lost, the shock wave must focus on the sources of the rings. 

Sugerman et al. \cite{Sugerman} observed by photon echoes, before the onset of the rings, a scattering \textquotedblleft hourglass\textquotedblright{} of denser matter. It seems that the rays that form the rings of SNR1987A are tangential (rather slightly secant) to the hourglass, into our direction, so that a long path in the excited gas close to the surface shows the rings.  Burrows et al.\cite{Burrows} criticized this model, showing that the variation of brightness close to the limb is too slow and concluding: \textquotedblleft A proper understanding of this system will require new physical insight\textquotedblright{}. Supposing that the hourglass is a distorted Str\"omgren shell, superradiance solves the problem.

The spectrum of the main ring is made of sharp emission lines, with Lyman $\alpha$ particularly intense. Light from a point on the ring propagates along a column of gas that can be cold, but excited by the hydrogen lines. This column of gas can be superradiant at its eigenfrequencies, in the same direction.

\subsection{Multiphoton and parametric interactions.}
In a Str\"omgren system, the temperatures of light beams have four orders of magnitude:

 {} i) The radial rays, emitted by the star have, at all frequencies, an initial temperature larger than $3.10^5 K$. 

 {} ii) The superradiant rays considered in the previous subsection are much less hot.

 {} iii) Spontaneously emitted and not notably amplified lines, in particular Lyman lines of atomic hydrogen have lower radiances.

 {} iv) The thermal background has the coldest temperatures.

The temperature of the radial rays allows multiphotonic interactions including several multiphoton transitions, even passing through virtual levels. Thus, frequencies may be combined to cool {\it at all frequencies} the temperature of these rays by a pumping of hydrogen atoms to high levels, including ionization levels.

The superradiant rays are hot enough to induce a strong de-excitation of the atoms. Pumping and de-excitations are combined into a parametric effect, a \textquotedblleft multiphotonic, parametric, induced scattering\textquotedblright{}  which transfers most of the energy of the continuous spectrum of radial, extremely hot rays to much cooler energy of the discrete spectrum of hydrogen in the superradiant rays. This  parametric multiphotonic scattering is strong one hand because it corresponds to a large increase of entropy, second because all involved rays are very hot.

The order of magnitude of temperature of radial rays decreases to the order of magnitude of the amplified superradiant rays. Suppose that the order of magnitude of the size of the star is an astronomical unit while the order of magnitude of the source of the superradiant beams is a fraction of light-year. In despite of the addition of all spectral radiances of the star, its solid angle of observation is so lower than the solid angle of observation of a superradiant dot, that the star is invisible.

\medskip
The standard theory of SNR1987A is unable to explain the total disappearance of the star. The coherent transfer of almost all the energy emitted by the star to the rings solves the problem of their persistent brightness.

\subsection{Impulsive Raman scattering.}
As the observation of this small effect requires large column densities of gas, the contributions of gases other than hydrogen seems negligible. Hydrogen must be in an excited state, in practice in the 2S or 2P states because atomic quadrupolar transitions frequencies in more excited states are low and thus provide a priori negligible effects.

This excited hydrogen may result from a thermal heating, from a Lyman $\alpha$ pumping of too cold (10 000-40 000K) atomic hydrogen, or from a cooling of an hydrogen plasma.
The result of the parametric effect catalyzed by the atomic excited hydrogen is frequently a transfer of energy from light which is redshifted to the background thermal radiation which is blueshifted, that is heated.

\subsubsection{Redshifts by thermally excited hydrogen.}4.3.1
The frequency shifts of a particular ray depends on energies transferred from or to all crossing beams, that is, assuming a statistical isotropy of the transfer, on their spectral irradiances. Far from a star, the irradiance of its rays is low, we can neglect their contribution to a blueshift.

\medskip
\begin{figure}[h]
\includegraphics[height=8 cm]{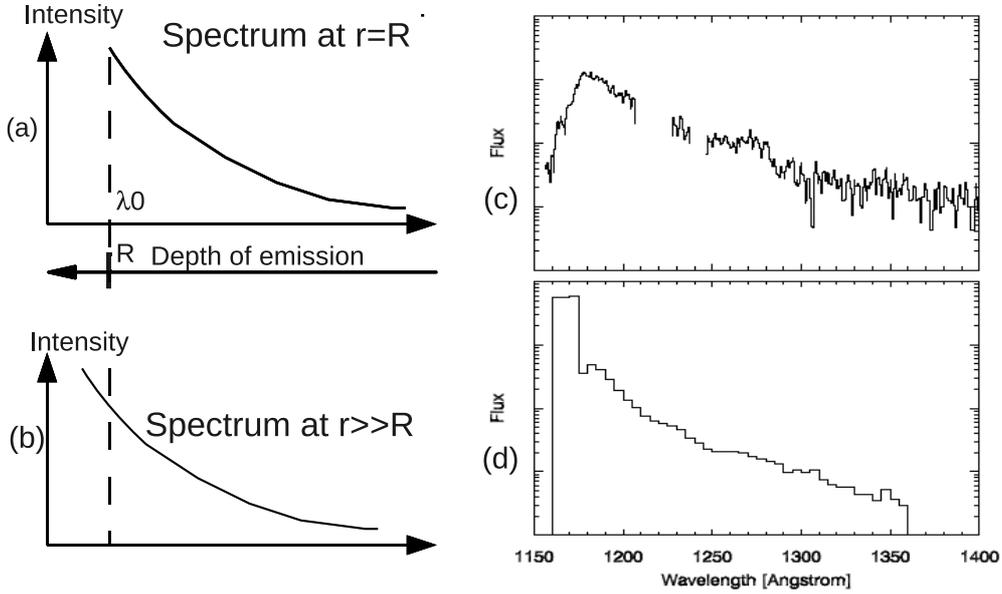}
\caption{Spectrum of weak light emitted inside the ring: (a), (b) present theory, observed at distances $R$ (a) and larger (b); (c) experimental and (d) theoretical from Michael et al. \cite{Michael}. $\lambda 0$ is Ly$_\alpha$ wavelength.}
\label{rad1}
\end{figure}

In the outer regions of a Str\"omgren sphere, it appears excited atoms whose increasing density has an exponential bearing until $r$ reaches value $R$. A spontaneous emission very close to the surface of the sphere is relatively intense, and light outputs the sphere without encountering atoms able to catalyze a redshift. Intensity emitted per unit volume decreases rapidly with depth $d=(R-r)$ and becomes negligible. The column density of excited atoms, thus the redshift of light, grows with $d$, so that the emission volume corresponding to a wavelength shift $\Delta\lambda$ grows as a function of $d$. The corresponding compensation of the variation of intensity is not sufficient to prevent its fall to zero on the spectrum (fig.\ref{rad1}a).

For $r$ slightly larger than $R$, a lot of energy is transferred from the radial beams to the tangential superradiant beams. Thus, the radial speed of propagation of radiant energy becomes very low, the irradiance of hot beams becomes very large. The cooler spontaneous beams, are blueshifted (fig.\ref{rad1}b), but along a short path because hydrogen is de-excited by the superradiant beams. 

The spectrum observed in the ring of SNR 1987A (fig.\ref{rad1}c) integrates spectra corresponding to various incidences, so that the break of intensity at short wavelengths is not sharp. It does not present a high spike which seems cut on the theoretical spectrum obtained from Monte-Carlo computations (fig.\ref{rad1}d).

The spectrum (fig.\ref{rad1}a) is similar to the spectrum of a Lyman break galaxy: these galaxies could be relatively close stars surrounded by clouds of hydrogen.

\subsubsection{Redshifts by hydrogen excited by Lyman $\alpha$ radiation:  Karlsson's constant.}

The very hot stars, and stars observed close to very hot objects have often anomalously large redshifts. Between 10 000 K and 40 000 K, hydrogen is hot enough for an atomic state, but too cold for a thermal excitation.

The propagation of far UV, X continuous spectrum light, in atomic hydrogen is remarkable:

 Light is absorbed at Ly$_\alpha$ frequency, 2P hydrogen is produced which catalyzes a transfer of energy to the background, producing a redshift. An equilibrium is reached in which absorption compensates various de-excitations of the atoms. There are two cases:

i) The column density of excited hydrogen resulting from Ly$_\alpha$ absorption along a path $\Delta x$ would shift the frequencies of light of more than the linewidth of Ly$_\alpha$ line. The absorption is limited by the redshift to a value $\Delta I$. As long as the initial radiance is larger or equal to $\Delta I$, the redshift is permanent. Absorptions and emissions of all lines of the gas follow the redshift. The linewidths corresponding to the redshift are large, so that the lines are not observable.

ii) If the radiance is lower than $\Delta I$, the equilibrium $T_{B\nu}=T_{P\nu}$ is reached without a redshift sufficient to shift an unabsorbed spectral element to Ly$_\alpha$ frequency. The redshift stops, all lines of the gas are visibly absorbed or emitted. If the low radiance results from a previous absorption of a line, a radiance larger than $\Delta I$ may be reached after a slow redshift resulting from the excitation of higher states, more probably from a decay of these states which populates the 2P state and, mainly, the metastable 2S state. Several beams may pump the atoms, so that $\Delta I$ becomes an irradiance. 

During a stop of the redshift (case ii), the Lyman $\beta$ and $ \gamma$ lines are absorbed making future \textquotedblleft previous absorptions\textquotedblright, so that the rays are multiplied with redshifts $(\nu_\beta-\nu_\alpha)/\nu_\alpha)=3*0,062$ and $(\nu_\gamma-\nu_\alpha)/\nu_\alpha)=4*0,062$. The remarkable constant 0,061 was observed par Burbidge and Karlsson \cite{Burbidge,Karlsson} in the Lyman forests of the quasars. 

The accreting neutron stars have never been observed. We searched the spectrum of a model consisting of a tiny, very hot source having a Str\"omgren shell so small that its pressure is of the order of 1 000 Pa. The spectrum shows, with decreasing frequency shifts, sharp, superradiant emission lines of the shell, large and saturated lines, then the Lyman forest \cite{MB03I,MB06A}: Some accreting neutron stars may be named quasars.

\medskip
At least a fraction of the \textquotedblleft cosmological redshift\textquotedblright{} results from a CREIL effect catalyzed by traces of atomic hydrogen excited by UV radiation in the interstellar and intergalactic regions. These regions are split into smaller regions where, depending on whether the UV irradiance is, before Ly$_\alpha$ absorption, higher or lower than $\Delta I$, light is very weakly or more strongly redshifted. Through the regions of high redshift, the distance to Earth is over-evaluated so that the maps of galaxies are popped: The inner part of a cellulose sponge is surprisingly similar to our maps of galaxies.

\subsubsection{Frequency shifts of the far UV emission lines of the Sun.}
Peter \& Judge \cite{Peter} have studied the frequency shifts of the UV lines emitted by the Sun, using spectra recorded by the Solar Ultraviolet Measurement of Emitted Radiation (SUMER) instrument on SOHO. They assume that vertical motions of the solar material produce these shifts by a Doppler effect. Thus, they assume that the shift is null at the rim of the solar disk, so that they use the observations at the rim to calibrate the spectrometer. Unhappily, they find wavelengths which differ from the laboratory values. The hottest line, Fe $_{XII}$ is blueshifted.

In despite of a high temperature, high pressure atomic hydrogen works as a crystal in which the atoms have sharp lines, so that a CREIL effect is possible. Suppose that the frequency shifts result mainly from CREIL.
The emission of the UV lines occurs mainly where the phonons have the energy of emitted quanta, deeper for shorter wavelength. A ray emitted at a certain frequency, then at a certain depth, crosses to the surface a longer distance at the limb than at the center of the disc: Thus the shift is larger at the limb than at the center. It is the opposite of the hypothesis of Peter \& Judge.

\medskip
CREIL effect explains the frequency shifts using laboratory frequencies\cite{MBSol}:

At the center of the disk: The shortest wavelength lines Fe$_{XII}$, Mg$_X$,..) lose energy along their whole path where they cross colder emissions, and background radiation; they are redshifted. 

Increasing the wavelength, the colder lines lose less energy along a shorter path, and may receive energy from Fe$_{XII}$ line, for instance, so that their redshift decreases down to zero.

Then the redshift is replaced by a blueshift; O$_V$, N$_V$ lines receive enough energy from the hot lines to be blue-shifted.

The frequency of the longest wavelength line (He$_I$) is not shifted because hydrogen atoms are not excited along its paths to the surface.

\subsubsection{ Microwaves frequency shifts in the solar wind.}
The solar wind cools while it expands, so that neutral, excited hydrogen atoms appear beyond 5 astronomical units.  A CREIL effect transfers energy from the solar light whose temperature is of the order of 5000 K and which is redshifted to colder beams:

When Pioneer 10 and 11 go beyond 5 AU, the very weak radio signals that they exchange with the Earth are blue-shifted (Anderson \& Laing \cite{Anderson}). If the increases of frequency are attributed to a Doppler effect, rather than to a CREIL effect, the probes appear to have an acceleration which can be interpreted only by changing the laws of gravitation, yet well proven. 

The very cold background is blue-shifted, that is heated \cite{MBSol}. This heating may explain that the low order terms of the spherical development of the cosmological background are bound to the ecliptic.

\section{Conclusion}

The geometrical propagation of light in homogeneous media depends only on the indices of refraction. The variation of the radiance of a monochromatic light beam  depends on an amplification coefficient whose sign is the sign of the difference $T_B-T_P$ between the Boltzman temperature of the involved transition and the Planck temperature of the beam. The usual Monte-Carlo computations which do not take the phases of the photon pilot waves into account, therefore ignore interferences, are wrong. They must be replaced by theories using coherent interactions initiated by Einstein and  developped for laser spectroscopy. Using the absolute radiance whose value is $h\nu^3/c^2$ in a blackbody at 0 K, Einstein coefficient A is null, so that the risk of phase or energy errors is lowered.

\medskip
Introducing the coherence in the study of elementary models leads to theoretical images and spectra very similar to observed images and spectra:

- The strong superradiant emission from the strongly emitting shell found by Str\"omgren produces dotted rings similar to the rings observed around SNR 1987A. The assumed high temperature light emission of the star allows a multiphotonic induced scattering which transfers almost all energy of the continuous spectrum of the star to the superradiant line spectrum, so that the star disappears.
For Einstein cross and other similar observations, superradiance of an amplifying shell may work more easily than gravitational lensings which require precise alignments of stars.

- A coherent parametric effect, made up of several coherent Raman effects, exchanges energy between beams of time-incoherent light in media obeying rules explained by G. L. Lamb, usually low pressure atomic hydrogen in 2S or 2P states. The frequency shifts resulting from these exchanges of energy explain many observations: Karlsson's periodicities in the spectra of the quasars,  \textquotedblleft anomalous acceleration\textquotedblright{} of the Pioneer probes and frequency shifts of the far UV lines of the Sun, preserving the laboratory measures of wavelengths.

\medskip
Monte Carlo method should be abandoned in astrophysics. New physics (Mond, variation of fundamental constants, \dots) should be introduced only after a serious testing of old, well verified physics. Accurate tests and developments of our proposals would be easily done by professional astrophysicists.


\begin{thebibliography}{}

\bibitem{Menzel}D. H. Menzel \textquotedblleft The dilution of radiation in a nebula\textquotedblright{} pasp {\bf 43}, 70-74 (1931). 
\bibitem{Einstein1917}A. Einstein, \textquotedblleft Zur Quantentheorie der Strahlung.\textquotedblright{} Phys. Z. {\bf 18}, 121-128 (1917). 
\bibitem{Planck1911}Planck M., \textquotedblleft Eine neue Strahlungshypothese.\textquotedblright{} Verh. Deutsch. Phys. Ges. {\bf 13}, 138-175 (1911).
\bibitem{Einstein1913}A. Einstein  \& O. Stern, \textquotedblleft Einige Argumente f\"ur die Annahme einer molekularen Agitation beim absoluten Nullpunkt.\textquotedblright{} Annalen der Physik {\bf 345}, 551-560 (1913)
\bibitem{Lamb}G. L. Lamb Jr., \textquotedblleft Analytical description of ultra-short optical pulse propagation in a resonant medium,\textquotedblright{}  Rev. Mod. Phys. {\bf 43}, 99-124 (1971).
\bibitem{Weiner}A. M. Weiner, D. E. Leaird, G. P. Wiederrecht, and K. A. Nelson, \textquotedblleft Femtosecond multiple-pulse impulsive stimulated Raman scattering spectroscopy,\textquotedblright{}  J. Opt. Soc. Am. {\bf B 8}, 1264-1275 (1991) 
\bibitem{MB}J. Moret-Bailly, \textquotedblleft Structure and light emission of a Str\"omgren system.\textquotedblright{} arXiv:0905.0554 (2009).
\bibitem{Stromgren}B. Str\"omgren, \textquotedblleft The physical state of interstellar hydrogen,\textquotedblright{} Astrophys. J., {\bf 89}, 526-547 (1939).
\bibitem{1987A}S. Immler, K. Weiler, R. McCray (Eds.)\textquotedblleft  Supernova 1987A: 20 Years After,\textquotedblright{} AIP Conf. Proc.,  {\bf 937}, ISBN: 978-0-7354-0448-9 (2007)
\bibitem{Chevalier}R. A. Chevalier \& V.V.Dwarkadas \textquotedblleft The presupernova H$_{II}$ region around SN 1987A,\textquotedblright{} ApJ, {\bf 452}, L45-L48 (1995).
\bibitem{Sugerman}B. E. K. Sugerman, A. P. S. Crotts, W. E. Kunkel, S. R. Heathcote, \& S. S. Lawrence, \textquotedblleft The three-dimensional circumstellar environment of SN 1987A.\textquotedblright{} arXiv: 0502378 (2005).
\bibitem{Burrows}C. J. Burrows, J. Krist, J. J. Hester,R. Sahai, J. T. Trauger, K. R. Stapelfeldt, J. S. Gallagher III, G. R. Ballester, S. Casertano, J. T. Clarke, D. Crisp, R.W. Evans, R. E. Griffiths, J. G. Hoessel, J. A. Holtzman, J. R. Mould, P. A. Scowen, A. M. Watson, \& J. A. Westphal, \textquotedblleft  Hubble space telescope observations on the SN 1987A triple ring nebula\textquotedblright{} ApJ, {\bf 452} 680-684 (1995)
\bibitem{Michael}E. Michael, R. McCray, R. Chevalier, A. V. Filippenko, P. Lundqvist, P. Challis, B. Sugerman, S. Lawrence, C. S. J. Pun,  P. Garnavich, R. Kirchner, A. Crotts, C. Fransson, W. Li, N. Panagia, M. Phillips, B. Schmidt, G. Sonneborn, N. Suntzeff, L. Wang, \& J. C. Wheeler,  \textquotedblleft Hubble space telescope observations of high-velocity Ly$\alpha$ and H$\alpha$ emission from supernova remnant 1987A: the structure and development of the reverse shock\textquotedblright{}, Astrophys. J., {\bf 593}, 809-830 (2003).
\bibitem{Burbidge}G.Burbidge, \textquotedblleft The distribution of redshifts in quasi-stellar objects, N-systems, and some radio and compact galaxies\textquotedblright{}, ApJ, {\bf 154,} L41-L48 (1968).
\bibitem{Karlsson}K. G. Karlsson, \textquotedblleft Possible discretization of quasar redshifts\textquotedblright, A\&A, {\bf 13,} 333-335 (1971).
\bibitem{MB06A}J. Moret-Bailly,  \textquotedblleft Propagation of light in low pressure ionized and atomic hydrogen: Application to astrophysics\textquotedblright, IEEETPS, {\bf 31}, 1215-1222 (2003).
\bibitem{MB03I}J. Moret-Bailly,  \textquotedblleft The parametric light-matter interactions in astrophysics\textquotedblright, AIP conf. proc., {\bf 822}, 226-238 (2006).

\bibitem{Peter}H. Peter \& P. G. Judge, \textquotedblleft On the Doppler shifts of solar ultraviolet emission lines.\textquotedblright{}, ApJ, {\bf 522}, 1148-1166 (1999).
\bibitem{MBSol}J. Moret-Bailly, \textquotedblleft Anomalous frequency shifts in the solar system\textquotedblright, arXiv:0507141 (2005).
\bibitem{Anderson}J. D. Anderson, P. A. Laing, E. L. Lau, A. S. Liu, M. M. Nieto, \& S. G. Turyshev, \textquotedblleft Study of the anomalous acceleration of Pioneer 10 and 11\textquotedblright, arXiv:0104064 (2001).
\end{thebibliography}
\end{document}